\documentclass[12pt]{article}
\usepackage{amsmath,bm,graphicx,epsfig}
\include{epsf} 
\newcommand\as{\alpha_S} 
\def\ltap{\raisebox{-.6ex}{\rlap{$\,\sim\,$}} \raisebox{.4ex}{$\,<\,$}} 
\def\gtap{\raisebox{-.6ex}{\rlap{$\,\sim\,$}} \raisebox{.4ex}{$\,>\,$}} 
\def\tL{{\tilde L}}
\def\nn{\nonumber} 
 
\def\HH{{H\!H}}

\begin{document}
\begin{titlepage}
\begin{flushright}
MPP-2016-152\\
TIF-UNIMI-2016-7
\end{flushright}

\renewcommand{\thefootnote}{\fnsymbol{footnote}}
\vspace*{2.5cm}

\begin{center}
{\Large \bf Transverse-momentum resummation \\
for Higgs boson pair production at the LHC  \\ with top-quark mass effects\\ }
\end{center}

\par \vspace{2mm}
\begin{center}
{\bf Giancarlo Ferrera${}^{(a)}$} ~and~ {\bf Jo\~{a}o Pires${}^{(b)}$}\\

\vspace{5mm}

${}^{(a)}$ Dipartimento di Fisica, Universit\`a di Milano and\\ INFN, Sezione di Milano,
I-20133 Milan, Italy\\\vspace{1mm}

${}^{(b)}$ Max-Planck-Institute for Physics,\\ F\"{o}hringer Ring 6, 80805 M\"{u}nchen, Germany\\

\end{center}

\vspace{1.5cm}


\par \vspace{2mm}
\begin{center} {\large \bf Abstract} \end{center}
\begin{quote}
\pretolerance 10000

We consider Higgs boson pair production via gluon fusion in hadronic collisions.
We report the calculation of the transverse-momentum ($q_T$) distribution of 
the Higgs boson pair with top-quark mass ($M_t$) effects fully taken into account. 
At small values of $q_T$ we resum the logarithmically-enhanced perturbative QCD contributions
up to next-to-leading logarithmic (NLL) accuracy. 
At intermediate and large values of $q_T$ we consistently combine resummation with the ${\cal O}(\as^3)$ fixed-order results. 
After integration over $q_T$, we recover the next-to-leading order (NLO) result for the inclusive cross section with full dependence on $M_t$. 
We present illustrative numerical results at LHC energies, together with an estimate of the corresponding perturbative uncertainties,
and we study the impact of the top-quark mass effects. 

\end{quote}

\vspace*{\fill}
\vspace*{2.5cm}

\begin{flushleft}
September 2016
\end{flushleft}
\end{titlepage}

\setcounter{footnote}{1}
\renewcommand{\thefootnote}{\fnsymbol{footnote}}

\section{Introduction}

Since the discovery of the Higgs boson ($H$)~\cite{Aad:2012tfa,Chatrchyan:2012xdj}
 during Run I of the Large Hadron Collider (LHC), it became a physics goal for Run II and future high-energy collider 
facilities to complete our understanding of the electroweak symmetry breaking mechanism of the Standard Model (SM). 
The study with increased precision of the couplings between the Higgs boson and the SM particles
shows a determination of the couplings to vector bosons and heavy fermions 
compatible with the SM values with an overall $15\%$ to $20\%$ uncertainty~\cite{HCMS,HATLAS,Hcombo}. 
The results obtained in Refs.~\cite{HCMS,HATLAS,Hcombo} concern single Higgs boson 
production.

In order to probe the Higgs boson self couplings one can consider the process of double Higgs boson ($\HH$)
production~\cite{Aad:2014yja,Khachatryan:2015yea,Aad:2015uka,Aad:2015xja,Khachatryan:2016sey,Aaboud:2016xco}. 
In this case, similarly to single Higgs boson production, the main production mechanism is  driven by gluon fusion.
At the leading order (LO), the two Higgs bosons can couple to a heavy-quark loop via a box diagram
or, via the trilinear Higgs self-coupling, to an off-shell Higgs boson produced by a triangular heavy-quark loop.
For this reason, the observation of Higgs boson pair production 
gives access to a direct extraction of the Higgs trilinear self-coupling and to the reconstruction of 
the Higgs boson potential~\cite{Dolan:2012rv,Baglio:2012np,Goertz:2013kp,Barger:2013jfa,deLima:2014dta,Dawson:2015oha}.

Predictions for double Higgs boson production in gluon fusion at LO were obtained in 
Refs.~\cite{Eboli:1987dy,LOHH1,LOHH2} including full top-quark mass ($M_{t}$) effects. 
However, since the gluon fusion mechanism for $\HH$ production is a loop induced process, the next-to-leading order (NLO) QCD corrections
were first obtained in the heavy top quark limit $M_{t}\to \infty$~\cite{NLOHH1} using the Higgs effective field theory (HEFT).
In this approximation the top-quark mass is regarded much larger than any other scale 
in the process and the top quark is integrated out at the Lagrangian level.
This significantly simplifies the calculation of the NLO corrections since the top-quark loops
shrink to a point-like interaction of the Higgs bosons with gluons. More recently, 
next-to-next-to-leading order (NNLO) 
predictions for $\HH$ production in the HEFT have been completed in Refs.~\cite{deFlorian:2013uza,deflorian-mazzitelli,Grigo:2014jma,NNLOHH1}.
Threshold resummation up to next-to-next-to-leading logarithmic (NNLL) accuracy in the HEFT has been performed in 
Refs.~\cite{Shao:2013bz} and~\cite{deFlorian:2015moa} matching the resummed results respectively with NLO and NNLO fixed-order calculations.

However, Higgs boson pairs are produced  with an invariant mass $(M_{\HH})$
which is above the top-quark mass threshold where the validity of the HEFT description breaks down.
Therefore, various approximations to include finite $M_t$ effects beyond LO have been performed in the literature.
In the so-called ``Born-improved HEFT'' approximation a reweighting of the NLO HEFT result is performed using a factor $B_{FT}/B_{HEFT}$, 
where $B_{FT}$ and $B_{HEFT}$ denote the LO matrix element squared in the full theory and in the HEFT respectively~\cite{NLOHH1}.
In the NLO calculation in Refs.~\cite{NLOmt1,NLOmt2} the top-quark mass dependence is fully taken into account in the real emission correction,
while the virtual amplitude is computed in the heavy top quark limit and reweighted by the Born-improved factor. 
HEFT results at NLO and NNLO improved by an expansion in $1/M_{t}^{2}$ have been obtained 
in Refs.~\cite{Grigo:2014jma,Grigo:2013rya,mtexp,Degrassi:2016vss}.

The NLO calculation including the full top-quark mass effects in both the real and virtual corrections  
has been performed only recently in Refs.~\cite{NLOHHfullmt,NLOHHfullmt2}. 
It shows that the total cross section is about 14\% smaller than the one obtained within the Born-improved HEFT approximation and that for  
values of the Higgs boson pair invariant mass 
beyond $M_{\HH} \sim 500$~GeV, the top-quark mass effects lead to a reduction of the differential cross section 
by about 20-30\% with respect to the same approximation.
Therefore, in order to get reliable predictions for the Higgs boson pair production cross section and corresponding distributions, 
it is important to include the full top-quark mass dependence. 

Among the various kinematical distributions, a particularly significant role is played by the transverse-momentum ($q_T$) spectrum of
the Higgs boson pair. A precise description of this observable is important to improve the statistical significance in the experimental searches
and therefore, it is essential to carefully investigate the theoretical uncertainties dominated by the higher-order QCD corrections.
In the large $q_T$ region ($q_T \sim M_{\HH}$)  
fixed-order calculations are theoretically justified. However, in the small $q_T$ region 
($q_T \ll M_{\HH}$) the reliability of the fixed-order perturbative expansion is spoiled by the presence of large logarithmic terms of the type 
$\alpha_S^n\log^m(M_{\HH}^2/q_{T}^2)$ which make the fixed-order results divergent in the limit $q_T \to 0$. 
In order to obtain reliable predictions 
at small $q_T$,
such large logarithmic contributions have to be 
systematically resummed to all orders~\cite{Dokshitzer:hw}--\cite{Catani:2013tia}.  
At intermediate values of $q_{T}$ the resummed and fixed order results can be consistently 
matched in order to get a uniform theoretical accuracy for the entire range of transverse momenta.

We have used the formalism introduced in Refs.~\cite{Catani:2000vq,Bozzi:2005wk} to perform the transverse momentum resummation for Higgs boson 
pair production up to next-to-leading logarithmic (NLL) accuracy, combining it with the NLO (i.e.\ ${\cal O}(\as^3)$) result
with full top-quark mass dependence.  
The implementation of our calculation for the $q_T$ spectrum was performed starting from the numerical code \texttt{HqT}~\cite{Bozzi:2005wk,Hqt2}.

The paper is organised as follows. In Sect.\ 2 we briefly review the resummation formalism of
Refs.~\cite{Catani:2000vq,Bozzi:2005wk,Catani:2013tia}. 
In Sect.\ 3 we present numerical fixed-order and resummed results for the transverse-momentum distribution of Higgs boson pairs
and we study the scale dependence of our results in order to estimate the perturbative uncertainty of our predictions. 
We also comment on the size of the finite top-quark mass effects. 
In Sect.\ 4 we present our conclusions.

\section{Transverse-momentum resummation}
\label{Sec:HHqt}
The resummation formalism used in this paper has been introduced in Refs.~\cite{Catani:2000vq,Bozzi:2005wk} 
and can be applied to a generic process where a high-mass system
of non strongly-interacting particles is produced in hadronic collisions. 
In this Section we briefly recall the main points of the formalism, by considering the specific case of the hadroproduction of
Higgs boson pairs in gluon fusion. For a detailed discussion we refer to 
Refs.~\cite{Catani:2000vq,Bozzi:2005wk,Bozzi:2007pn,Catani:2010pd,Catani:2013tia}.

The transverse-momentum differential cross section for this process can be written as\,\footnote{In this Section we denote with 
$d\hat\sigma_{\HH}/{d q_T^2}$ the double differential cross section $M^2 d\hat\sigma_{\HH}/{d M^2 d q_T^2}\,$.}:
\begin{equation}
\!\!\!\frac{d\sigma_{\HH}}{d q_T^2}(q_T,M,s)=\sum_{a_1,a_2} \int_0^1 \!dx_1 \int_0^1 \!dx_2 f_{a_1/h_1}(x_1,\mu_F^2) f_{a_2/h_2}(x_2,\mu_F^2)
\frac{d{\hat \sigma}_{\HH\,a_1a_2}}{d q_T^2}(q_T, M,{\hat s}; \as(\mu_R^2),\mu_R^2,\mu_F^2) 
\,,\!\!
\end{equation}
where $f_{a/h}(x,\mu_F^2)$ ($a=g,q,{\bar q}$) are the parton densities of 
the colliding hadrons ($h_1$ and $h_2$),
$d{\hat \sigma}_{\HH\,a_1a_2}/d q_T^2$ are the partonic cross sections, $M= M_\HH$ is the invariant mass of the Higgs boson pair, 
$s$ (${\hat s}=x_1x_2s$) is the hadronic  (partonic) centre-of-mass energy, $\mu_R$ and $\mu_F$ are respectively 
the renormalisation and  factorisation scale.

The partonic cross section is decomposed as follows:
${d{\hat \sigma}_{\HH\,a_1a_2}}=
{d{\hat \sigma}_{\HH\,a_1a_2}^{(\rm res.)}}
+{d{\hat \sigma}_{\HH\,a_1a_2}^{(\rm fin.)}}\,$.
The `resummed' component, ${d{\hat \sigma}_{\HH\,a_1a_2}^{(\rm res.)}}$, 
contains all the logarithmi\-cally-enhanced contributions at small $q_T$
which have to be evaluated to all orders in $\as$ and the `finite' component, ${d{\hat \sigma}_{\HH\,a_1a_2}^{(\rm fin.)}}$,
is free of such contributions. 

The resummation procedure is carried out in the impact-parameter ($b$) space. 
The resummed component is obtained by performing the inverse Bessel transformation with
respect to the impact parameter:
\begin{equation}
\frac{d{\hat \sigma}_{\HH \,a_1a_2}^{(\rm res.)}}{dq_T^2}(q_T,M,{\hat s};
\as(\mu_R^2),\mu_R^2,\mu_F^2) 
=\frac{M^2}{\hat s} \;
\int_0^\infty db \; \frac{b}{2} \;J_0(b q_T) 
\;{\cal W}_{a_1a_2}^{\HH}(b,M,{\hat s};\as(\mu_R^2),\mu_R^2,\mu_F^2) \;,
\end{equation}
where $J_0(x)$ is the $0$th-order Bessel function. 
The resummation structure of ${\cal W}_{a_1a_2, \,N}^\HH$ can be factorised and
organised in exponential form by considering the Mellin $N$-moments ${\cal W}_N$ of ${\cal W}$ 
with respect to  $z=M^2/{\hat s}$ at fixed $M$\,\footnote{Here, to simplify the notation, 
flavour indices are understood.}:
\begin{align}
{\cal W}_{N}^{\HH}(b,M;\as(\mu_R^2),\mu_R^2,\mu_F^2)
&={\cal H}_{N}^{\HH}\left(M, 
\as(\mu_R^2);M^2/\mu^2_R,M^2/\mu^2_F,M^2/Q^2
\right) \nonumber \\
&\times \exp\{{\cal G}_{N}(\as(\mu^2_R),\tL;M^2/\mu^2_R,M^2/Q^2
)\}
\;\;,
\end{align}
where we have defined the logarithmic expansion parameter $\tL= \ln \left({Q^2 b^2}/{b_0^2}+1\right)$, 
and  $b_0=2e^{-\gamma_E}$ ($\gamma_E=0.5772...$  is the Euler number).
The resummation scale $Q$ \cite{Bozzi:2005wk} 
parameterises the arbitrariness in the 
separation (factorisation) between finite and logarithmically-enhanced terms.
Variations of $Q$ around the hard scale $M$ can be used to estimate the effect of uncalculated higher-order logarithmic contributions.

The {\itshape universal} (process independent) form factor $\exp\{{\cal G}_N\}$
includes all the large logarithmic terms $\as^n\tL^m$, with $1 \leq m \leq 2n$
that order-by-order in $\as$  are logarithmically divergent as $b\to\infty$.
The exponent ${\cal G}_N$ 
can systematically be expanded in powers of $\as\equiv\as(\mu^2_R)$ 
as follows:
\begin{align}
\label{exponent}
{\cal G}_{N}(\as, \tL;M^2/\mu^2_R,M^2/Q^2)&=\tL \;g^{(1)}(\as \tL)+g_N^{(2)}(\as \tL;M^2/\mu_R^2,M^2/Q^2)\nn\\
&+\frac{\as}{\pi} g_N^{(3)}(\as \tL;M^2/\mu_R^2,M^2/Q^2)+\dots\;,
\end{align}
where the term $\tL\, g^{(1)}$ collects the leading logarithmic (LL) 
contributions, the function $g_N^{(2)}$ includes
the NLL contributions, $g_N^{(3)}$ controls the NNLL terms and so forth.
The logarithmic variable $\tL$ is equivalent to $L=\ln \left({Q^2 b^2}/{b_0^2}\right)$
when $Qb\gg 1$ (i.e.\ small values of $q_T$), but it leads to a behaviour
of the form factor at small values of $b$ such that $\tL\to 0$ and $\exp\{{\cal G}_N\}\to 1$  when $Qb\ll 1$.
The logarithmic expansion with respect to $\tL$ thus reduces the impact of large and 
unjustified resummed  contributions in the small-$b$ region (i.e.\ at large values of $q_T$), and it acts as
a perturbative unitarity constraint since
it allows us to exactly recover the fixed-order value
of the total cross section 
upon integration over $q_T$.

The hard-collinear function ${\cal H}_N^{\HH}$ 
fully encodes the process dependence of the resummation factor ${\cal W}_{N}^\HH$
and it includes all the perturbative terms that behave as constants in the limit $b\to\infty$. 
It has a customary perturbative expansion: 
\begin{align}
\label{hexpan}
{\cal H}_N^{\HH}(M,\as;M^2/\mu^2_R,M^2/\mu^2_F,M^2/Q^2)&=
\sigma_{\HH}^{(0)}(M)
\Bigl[ 1+ \frac{\as}{\pi} \,{\cal H}_N^{\HH \,(1)}(M^2/\mu^2_F,M^2/Q^2) ~
\Bigr. \nn \\
&+ \Bigl.
\left(\frac{\as}{\pi}\right)^2 
\,{\cal H}_N^{\HH \,(2)}(M^2/\mu^2_R,M^2/\mu^2_F,M^2/Q^2)+\dots \Bigr] \;\;,
\end{align}
where $\sigma_{\HH}^{(0)}$ is the Born-level partonic cross section for 
the process $g g \to \HH$.

The general structure of the hard-collinear function ${\cal H}_N^{F}$ has been obtained in Ref.~\cite{Catani:2013tia}, where
it is shown that the process dependent contribution to ${\cal H}_N^{F}$ can be embodied in
a single perturbative hard factor which is directly related to the finite part of the virtual amplitude of the corresponding process. 
The process independent part of the hard-collinear function ${\cal H}_N^{F}$ has been explicitly computed up to NNLO in 
Refs.~\cite{Catani:2011kr,Catani:2012qa}.
For $\HH$ production the NLO corrections in the full theory were recently calculated~\cite{NLOHHfullmt,NLOHHfullmt2}.
From the values of the virtual amplitude with full top-quark mass dependence computed in Refs.~\cite{NLOHHfullmt,NLOHHfullmt2} we 
extracted numerically the process dependent contribution to the NLO coefficient ${\cal H}_N^{\HH\,(1)}$. 

We now consider the finite component of the cross section. 
Since it does not contain large logarithmic terms, it 
can be computed at fixed order in perturbation theory
starting from the standard fixed-order results 
and subtracting the expansion of the resummed component  
at the same perturbative order~\cite{Bozzi:2005wk}.

In summary, the resummation at NLL+NLO
accuracy is obtained by including  the functions $g^{(1)}$, $g^{(2)}_N$
and the coefficient ${\cal H}_N^{\HH(1)}$ 
in the resummed component, and by computing the finite 
component at first order (i.e.\ at ${\cal O}(\as^3)$)\,\footnote{This matching procedure coincides with that of 
Refs.~\cite{Bozzi:2005wk,Bozzi:2010xn}. We note however that here we are using different labels.
The fixed-order label NLO used here directly refers to the perturbative accuracy in the small-$q_T$ region 
and of the total cross section, while the labels LO and NLO used in Refs.~\cite{Bozzi:2005wk,Bozzi:2010xn} 
refer to the perturbative accuracy in the large-$q_T$ region.}.
We note that the NLL+NLO
result includes the {\em full} NLO perturbative contribution in the small-$q_T$ region
and that the NLO result for the total cross section  
is exactly recovered upon integration
over $q_T$ of the differential cross section $d \sigma/dq_T$ at NLL+NLO
accuracy. 

\section{Numerical results for $\HH$ production at the LHC}

In this Section we consider Higgs pair production via gluon fusion in $pp$ collisions at the centre-of-mass energy of $\sqrt{s}=14$~TeV. 
We first show the fixed-order results which are valid in the large $q_T$ region and then 
we present our resummed prediction at NLL+NLO focusing on the small $q_T$ region. We include the full dependence on the top-quark mass 
and we comment on the size of the $M_t$ effects.

The hadronic cross section is computed using the PDF4LHC15 NLO parton densities~\cite{pdf1} 
with $\as$ evaluated at 2-loop order and $\as(m_Z^2)=0.118$ and
we consider $N_f = 5$ flavours of light quarks in the massless approximation.
We set the central value of the renormalisation, factorisation and resummation scales at $\mu_R=\mu_F=Q=M_{\HH}/2$.
The Higgs boson and top-quark masses have been set to $M_H=125$~GeV and $M_t = 173$~GeV respectively,
and the top quark and Higgs boson widths have been set to zero.

Bottom-quark mass effects in the double Higgs boson total cross section 
contribute well below $1\%$ level and have been neglected in the present study. 
We thus have a two-scale problem with $q_T \ll M$, where $M$ is the 
{\itshape hard}  scale  of the process $M\sim M_{HH}$
and the top-quark mass is of the same order of the hard scale. 
This fact justifies the application of the standard $q_T$ resummation
formalism to compute the $\HH$ $q_T$ spectrum with finite top-quark 
mass effects\,\footnote{We note that the case of single 
Higgs boson production is different, since bottom-quark mass effects 
are sizeable and their inclusion leads
to a three-scale problem~\cite{Grazzini:2013mca}.}.

We start the presentation of our numerical results by considering the calculation at fixed order.
As explained in the previous Section we performed the calculation of the double Higgs boson $q_T$ spectrum at first order in QCD (i.e.\ ${\cal O}(\as^3)$).
The relevant partonic subprocesses are $gg\to \HH g$, $qg\to \HH q$, $\bar{q}g\to \HH\bar{q}$ and $q\bar{q}\to \HH g$ and
we have generated all the relevant one-loop amplitudes using \texttt{GoSam}~\cite{gosam1,gosam2} retaining the full top-quark mass dependence. 
The corresponding matrix elements in the HEFT have been computed analytically. The phase space integration was performed
using the \texttt{CUBA} library~\cite{cuba1}.

\begin{figure}[t]
\begin{center}
\begin{tabular}{cc}
\includegraphics[width=0.46\textwidth]{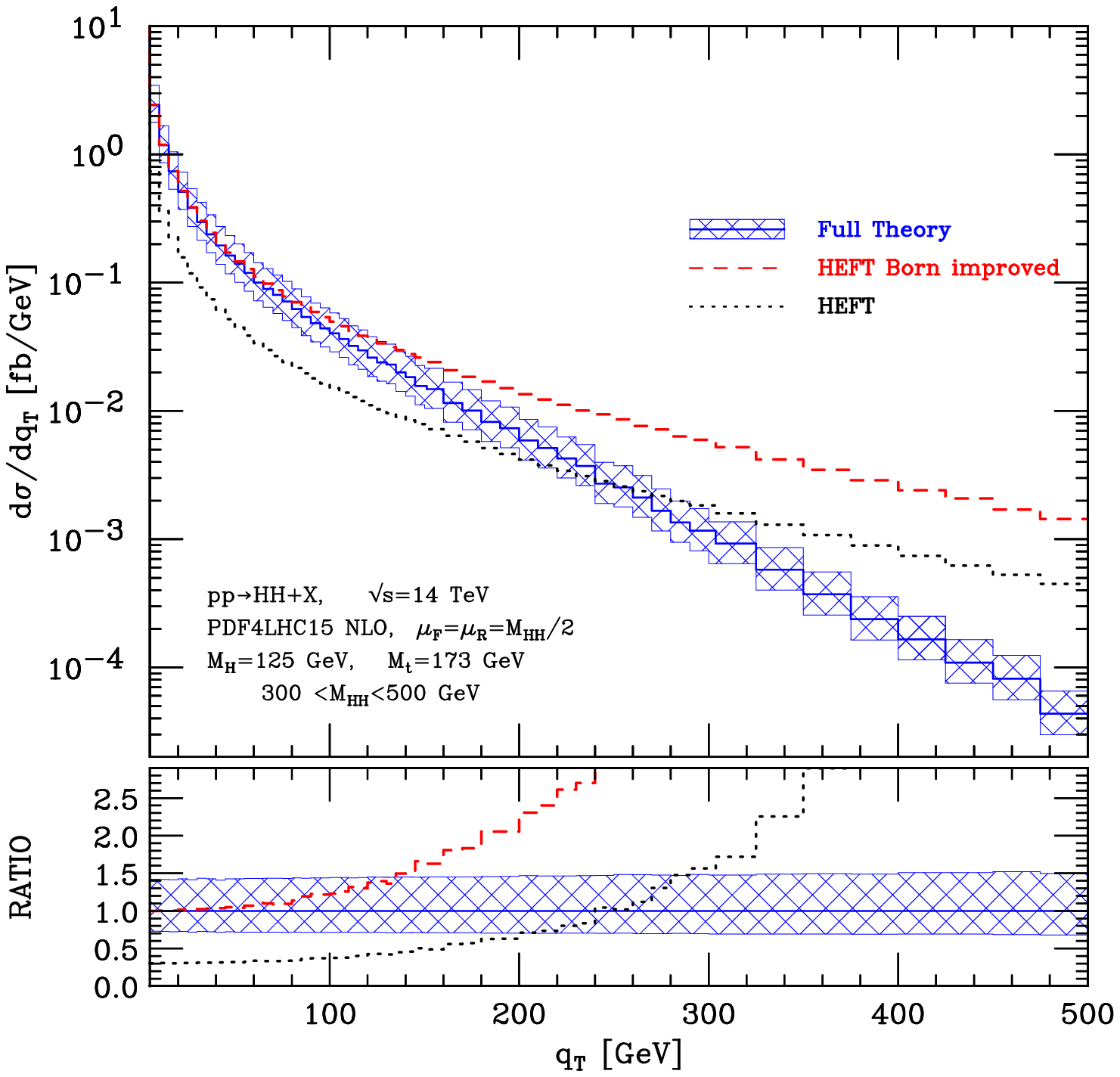} & \includegraphics[width=0.46\textwidth]{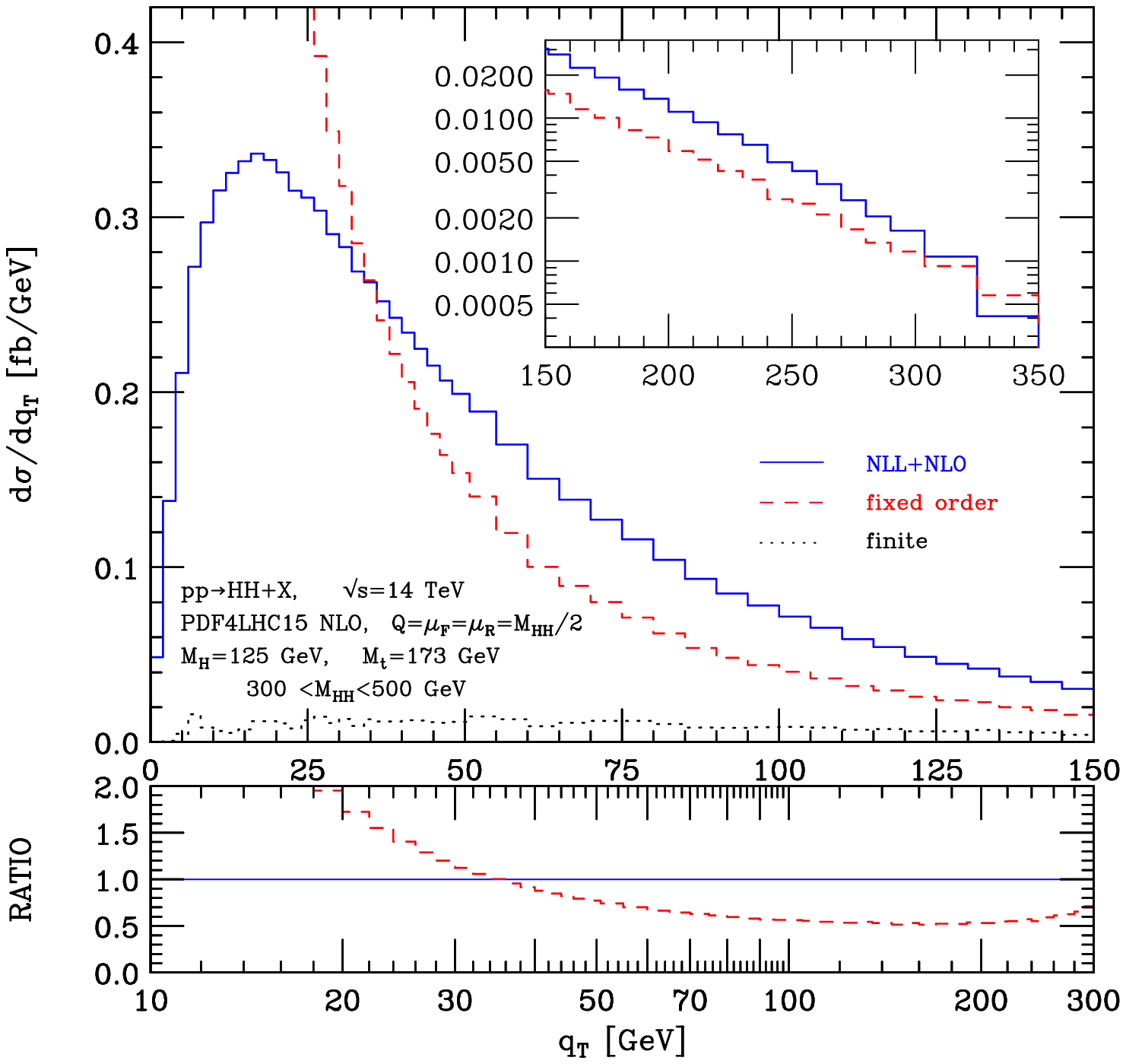}\\
\end{tabular}
\end{center}
\caption{\label{fig1}
{\em 
The $q_T$ spectrum of Higgs boson pairs at the LHC ($\sqrt{s}=14$~TeV).
Left panel: fixed-order prediction at ${\cal O}(\as^3)$ accuracy in the full theory (blue solid),
HEFT (black dotted) and Born-improved reweighted HEFT (red dashed).
The band is obtained by varying $\mu_R$ and $\mu_F$ as described in the text. 
Right panel: resummed prediction at NLL+NLO accuracy in the full theory. The resummed result (blue solid) is compared to the corresponding 
fixed-order result (red dashed) and to the finite component (black dotted).
The lower panels show the ratios with respect to the central value of the full theory result.
}}
\end{figure}
In Fig.~\ref{fig1} (left panel) we present the double Higgs boson $q_T$ spectrum for an invariant mass in the range $300 < M_{\HH}<500$~GeV 
at the LHC ($\sqrt{s}=14$~TeV).
We show the first order prediction in the full theory which includes the exact top-quark mass dependence (blue solid line)
and in the HEFT (black dotted).
In addition, we present also the approximation obtained by reweighting the HEFT result
by the Born-level matrix elements for $\HH$ production with the full top-quark mass dependence (red dashed). 
The reweighting is performed at the matrix element level using the initial-initial antenna phase space mapping~\cite{IImap}
to generate a Higgs boson pair Born-like configuration from the real-emission kinematics.

We show the scale dependence band (blue solid) of the full theory result which is obtained by varying independently the renormalisation and factorisation scales 
by a factor 2 around their central value, with the constraint $1/2\leq \mu_R/\mu_F\leq 2$.
The scale dependence band is about $\pm 35\%$ at small $q_T$, and it slightly increases up to $\pm 40\%$ at $q_T\sim 400-500$~GeV. We observe that the
band is rather large and flat. This is not unexpected since the bulk of the scale dependence is due to the $\mu_R$ dependence which is driven 
by the overall factor $\alpha_S(\mu_R)^3$.

By comparing the effective theory and the full theory results we observe that the pure HEFT calculation gives a poor approximation 
of the full theory spectrum for the entire range of $q_T$ (the two predictions accidentally cross each other for $q_T \sim 250$~GeV). 
The Born-improved reweighted HEFT result
gives a good approximation of the exact calculation in the small $q_T$ region ($q_T\ltap 50$~GeV) where however both results diverge
logarithmically. This is expected since in the small $q_T$ limit the phase space is
restricted to soft and collinear emissions and in this limit the fixed-order cross section factorises into the Born contribution
and process independent logarithmic terms. 
At larger $q_T$ ($q_T\gtap 50$~GeV), the agreement between the Born-improved HEFT and the full theory result rapidly deteriorates.
The top-quark mass effects in the loop diagrams produce deviations in the fixed order $q_{T}$ spectrum 
of about $20-25\%$ at $q_{T}\sim 100$~GeV and $80-100\%$ at $q_{T}\sim 175$~GeV. The $q_{T}$ spectrum in the Born-improved HEFT  approximation
is much harder than in the full theory and generates an unphysical tail at large $q_T$. 

We now turn to present the resummed results. In Fig.~\ref{fig1} (right panel) 
we compare the NLL+NLO spectrum (blue solid line) at the default scales ($\mu_F=\mu_R=Q=M_{\HH}/2$)
with the fixed-order result (red dashed). The finite component is also shown for comparison (black dotted). 
In the inset plot of the figure it is shown the region from intermediate to large values of $q_T$. We observe that while the fixed-order calculation diverges at
$q_T\to 0$, the resummation leads to a well-behaved distribution: it vanishes as $q_T\to 0$, has a kinematical peak at $q_T\sim 18$~GeV and tends to the
corresponding fixed-order result for $q_T\sim M_{\HH}$. The finite component vanishes as $q_T\to 0$ and  gives  a contribution to the NLL+NLO result that is around  
$4\%$ in the peak region and it increases to about $15\%$ at $q_T\sim 125$~GeV and $30\%$ at $q_T\sim 200$~GeV. 
We notice that in a wide region of intermediate values of $q_T$  the difference between
the NLL+NLO and the fixed-order result is quite large (around $40-50\%$ for $80\ltap q_T \ltap 250$~GeV), 
thus indicating that the effect of the logarithmic terms included in the resummation is important even outside the small-$q_T$
region. The contribution of the finite component sizeably increases at large values of $q_T$ ($q_T\sim M_{\HH}$) and the resummed spectrum approaches the
fixed order prediction. We have checked the numerical accuracy of our calculation by computing the integral over $q_T$ of the NLL+NLO resummed spectrum. 
The result is in agreement with the value of the NLO total cross section calculated in Refs.~\cite{NLOHHfullmt,NLOHHfullmt2} at the percent level,
thus proving that the uncertainty associated to the numerical extraction of the ${\cal H}_N^{\HH\,(1)}$ coefficient is completely under control.

\begin{figure}[t]
\begin{center}
\begin{tabular}{cc}
\includegraphics[width=0.46\textwidth]{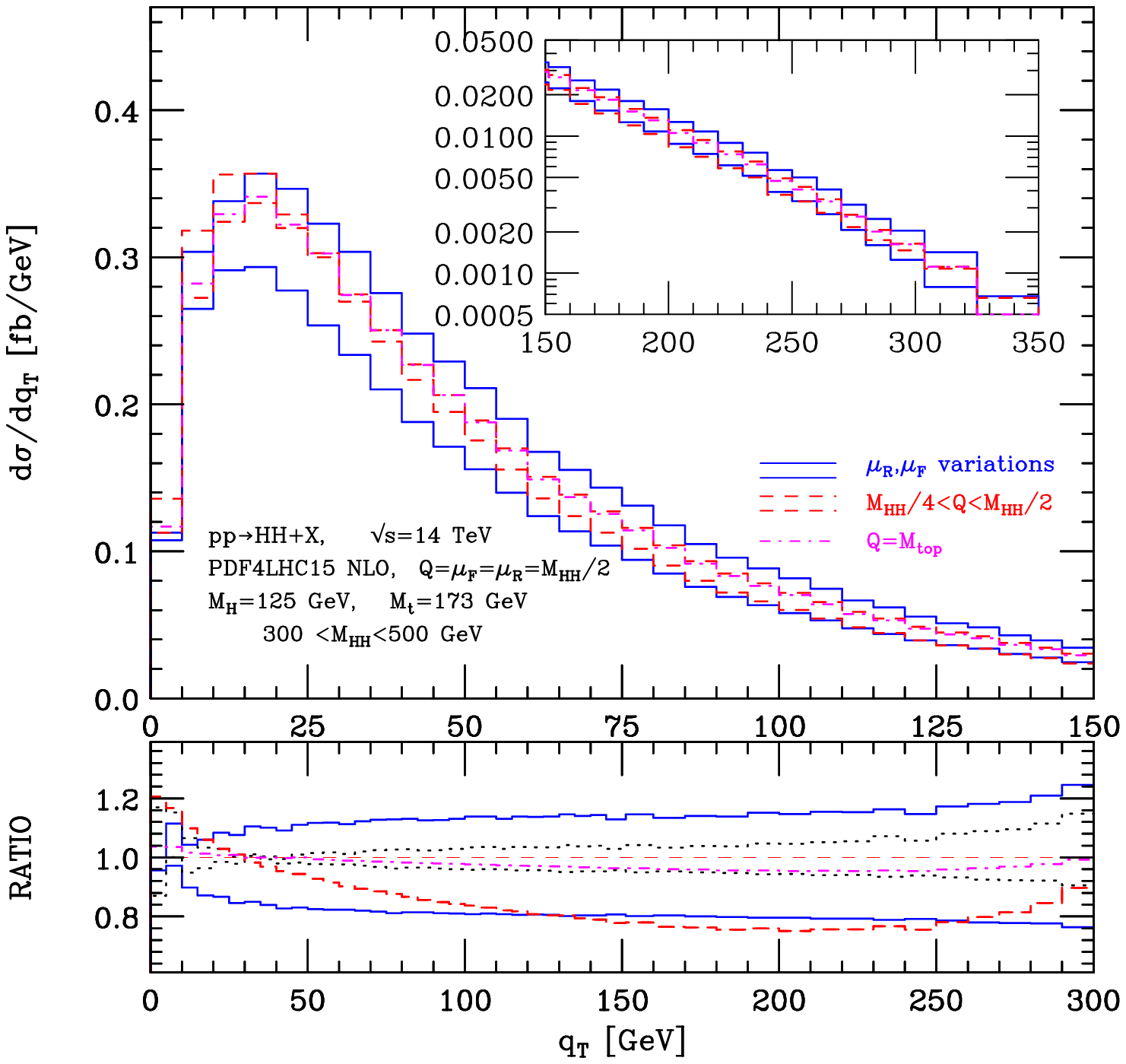} & \includegraphics[width=0.46\textwidth]{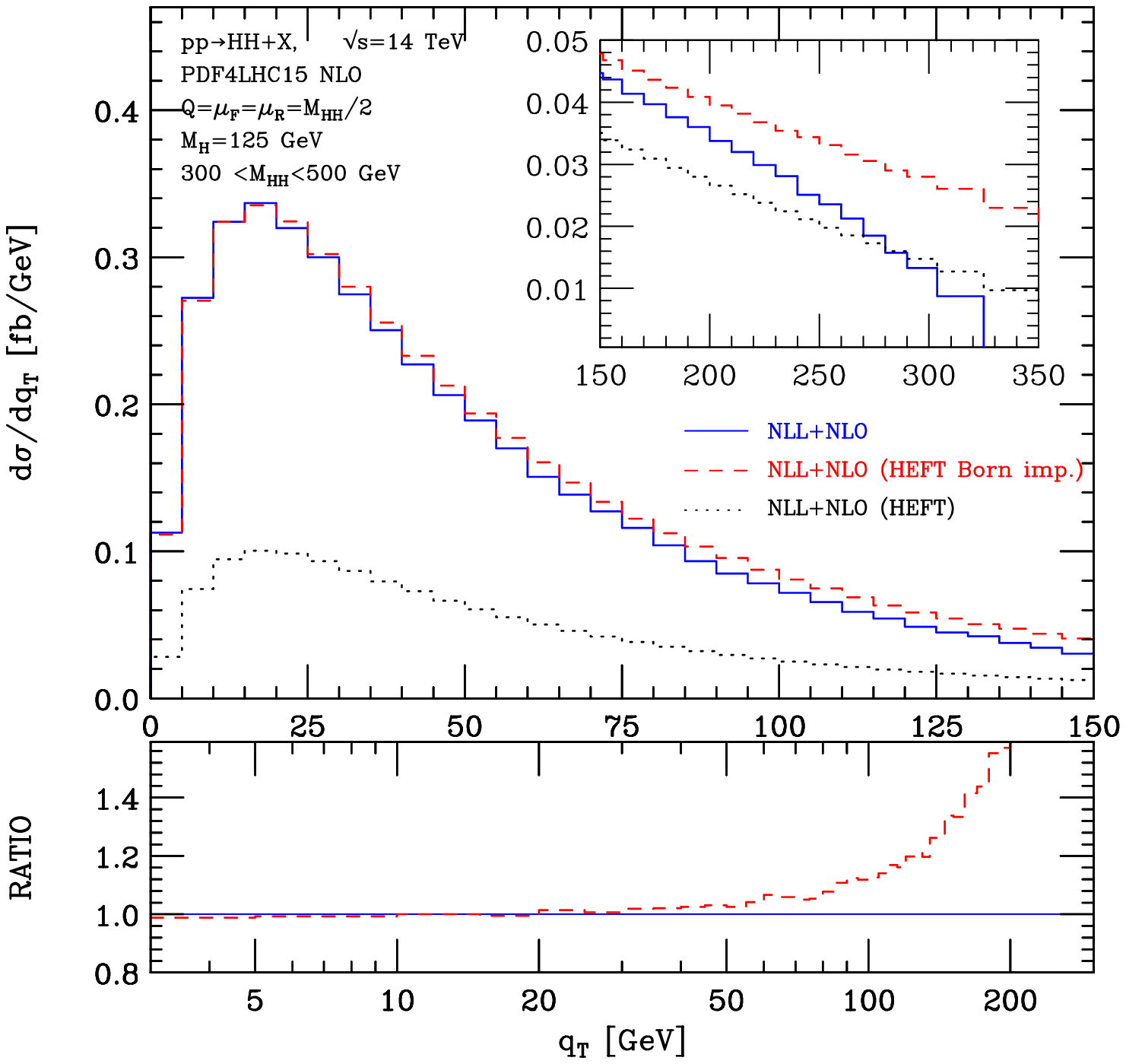}\\
\end{tabular}
\end{center}
\caption{\label{fig2}
{\em 
The $q_T$ spectrum of Higgs boson pairs at the LHC ($\sqrt{s}=14$~TeV).
Left panel: scale variation bands for the NLL+NLO result in the full theory.
The bands are obtained by varying $\mu_R$ and $\mu_F$ and $Q$ as described in the text. 
Right panel: resummed prediction at NLL+NLO accuracy in the full theory  (blue solid),
HEFT (black dotted) and Born-improved reweighted HEFT (red dashed).
The lower panels show the ratios with respect to the central value of the full theory result.
}}
\end{figure}
We now discuss the scale dependence of the NLL+NLO result. 
As previously discussed the resummation formalism we are using is strictly valid for a 
two-scale problem with the resummation scale of the order of the top-quark mass. 
For this reason we explicitly avoided values of the
resummation scale parametrically too large with respect to the top-quark mass $M_t$. 
In Fig.~\ref{fig2} (left panel)  we show 
the resummation scale dependence band (red dashed lines)  obtained 
by varying $Q$ in the region
$M_{\HH}/4\leq Q \leq M_{\HH}/2$ at fixed values of $\mu_R$ and $\mu_F$ ($\mu_R=\mu_F=M_{\HH}/2$).
The resummation scale dependence is about $\pm 3\%$ at the peak, decreases to about $\pm 1.5\%$ at $q_T\sim 30$~GeV  and increases again to
about $\pm 12\%$ at $q_T\sim 200$~GeV. 

Additionally we show in Fig.~\ref{fig2} (left panel) the resummed prediction for the resummation scale choice $Q=M_{top}$ (magenta dot-dashed line)
and we quantitatively estimated the effect of this choice. We observe no significant differences with respect to the choice $Q=M_{HH}/2$. 
The percentual difference with respect to our default value ($Q=M_{HH}/2$) is around $1\%$ at the peak, 
it decreases to few permille  at $q_T \sim 30$\,GeV, it increases to $2\%$ at $q_T \sim 100$\,GeV and it remains
$\ltap  4\%$ for $100 \ltap  q_T \ltap 300$\,GeV. This quantitative effect is widely covered by the resummation scale uncertainty band and 
can be explained by the fact that the $\HH$ cross section is peaked at an invariant mass of the order of $M_{HH} \simeq 400$\,GeV 
(for which $M_{HH}/2 \approx M_{top}$)~\cite{NLOHHfullmt,NLOHHfullmt2}. 

%

In Fig.~\ref{fig2} (left panel) we also considered
the renormalisation and factorisation scale dependence band (blue solid)  obtained by varying independently  $\mu_R$ and $\mu_F$ 
by a factor 2 around their central value (with the constraint $1/2\leq \mu_R/\mu_F\leq 2$) at fixed value of the resummation scale ($Q=M_{\HH}/2$).
The $\mu_R$ and $\mu_F$ scale dependence band is about $\pm 10\%$ at the peak and it increases to about $\pm 12\%$ at $q_T\sim 30$~GeV, 
to about $\pm 17\%$ at $q_T\sim 100$~GeV and to about $\pm 20\%$ at $q_T\gtap 250$~GeV.
We observe that the size of the $\mu_R$ and $\mu_F$ band is larger than the $Q$ band for a wide region of $q_T$ ($q_T\ltap 250$~GeV).
By comparing the $\mu_R$ and $\mu_F$ scale dependence bands of fixed order and resummed calculations, we observe that 
the resummed scale dependence band is not flat and its size is smaller than the fixed-order one at small and intermediate values of $q_T$ ($q_T\ltap 250$~GeV). 
This behaviour is not unexpected since the NLL+NLO resummed result, contrary to the fixed-order case, 
includes the full NLO correction in the small-$q_T$ region and satisfies the NLO unitarity constraint described at the end of Section~\ref{Sec:HHqt}
(see the discussion after Eq.~\eqref{exponent}) and these NLO effects are spread on a region from small to intermediate values of $q_T$. 
Nevertheless we point out  that the $\mu_R$ and $\mu_F$ scale dependence is only a part of the 
perturbative uncertainty of the resummed prediction which includes also the resummation scale dependence.

We add a comment on the relation between the scale variation bands and the normalisation of the resummed 
$q_T$ spectra which is given by the corresponding NLO total cross section. On the one hand, the total cross section
does not depend on the resummation scale. For this reason, the corresponding uncertainty band is 
independent on normalisation effects. On the other hand, the $\mu_R$ and $\mu_F$ scale
variation band depends on normalisation effects 
and can be substantially reduced if we consider the {\itshape normalised} $q_T$ spectrum, $1/\sigma \times d\sigma/dq_T$ (i.e.\ if we are interested only on the 
{\itshape shape}  of the $q_T$ distribution  and not on its normalisation). 
The $\mu_R$ and $\mu_F$  scale dependence band for the normalised $q_T$ spectrum is shown in the lower left panel in Fig.~\ref{fig2} (black dotted lines),
and we observe that it becomes smaller than the resummation scale uncertainty band for the entire $q_T$ range.

We conclude this Section with an assessment of the size of the finite top-quark mass effects which are included in our calculation.
In order to study the impact of the $M_t$ effects we computed the resummed spectrum also in the HEFT approximation
and in the Born-improved HEFT (the latter approximation was obtained by reweighting the HEFT result with the Born-improved factor
at a differential level)\,\footnote{We stress that the full theory result contains the complete finite $M_t$ effects both in the resummed part, 
through the Born-level partonic cross section $\sigma_{\HH}^{(0)}$ and the NLO coefficient ${\cal H}_N^{\HH\,(1)}$ (see Eq.~(\ref{hexpan})), 
and in the finite component.}.
In Fig.~\ref{fig2} (right panel) we compare the NLL+NLO prediction in the full theory with exact $M_t$ dependence
(blue solid line), with the pure HEFT (black dotted) and with the Born-improved HEFT (red dashed) results.
We observe, similarly to the fixed-order case, that the Born-improved HEFT gives a good approximation  (within $5\%$ accuracy) 
of the full theory result  for $q_T\ltap 70$~GeV\,\footnote{We note that the Born-improved HEFT approximation works particularly 
well (within $1\%$ accuracy) for $q_T$ values around the peak. This agreement is not general and it 
depends on the particular Higgs boson pair invariant mass window. 
Considering Higgs boson pairs with an invariant mass in the range $350 < M_{\HH}<400$~GeV the agreement in the peak region is about $7\%$.}. 
At higher values of $q_T$ we observe that the finite top mass effects are large and have a strong $q_T$ dependence. 
The effect is about $12\%$ at $q_{T}\sim 100$~GeV, about $60\%$ at $q_{T}\sim 200$~GeV and larger than $200\%$ for 
$q_{T}\gtap 250$~GeV, showing that the inclusion of the full top-quark mass dependence is essential to obtain a reliable description 
of the double Higgs boson $q_T$ spectrum over a wide region of $q_T$.

\section{Conclusions}

We have considered Higgs boson pairs produced in gluon fusion in hadronic collisions and
we performed the calculation of the transverse-momentum ($q_T$) distribution 
of the double Higgs boson system taking into account finite top-quark mass ($M_t$) effects.

At small values of $q_T$ we have resummed the logarithmically-enhanced perturbative QCD contributions
using the formalism introduced in Refs.~\cite{Catani:2000vq,Bozzi:2005wk}.
We have presented the results of the resummed calculation at next-to-leading logarithmic accuracy (NLL), and 
we have combined them with the fixed-order computation at ${\cal O}(\as^3)$. 
Our calculation includes the complete next-to-leading order (NLO) contributions at small $q_T$ 
and exactly reproduces the NLO total cross section with the full top-quark mass dependence upon integration over $q_T$.

We have presented illustrative numerical results in $pp$ collisions at $\sqrt{s}=14$~TeV, 
performing a study of the scale dependence of our predictions to estimate the corresponding perturbative uncertainty.
Comparing the NLL+NLO and fixed-order results, we have shown
that the higher-order terms contained in the resummed calculation are essential to obtain reliable predictions at 
small $q_T$ and give an important contribution ($\gtap 40-50\%$)
to the fixed-order result, in a wide region of intermediate values of $q_T$ ($q_T\ltap 250$~GeV).
Finally, by comparing our results with the  Born-improved  Higgs effective field theory (HEFT)
approximation in the $M_{t}\to \infty$ limit, we have quantified the size of the finite $M_t$ effects 
which turn out to be large ($\gtap 60\%$) for $q_T\gtap 200$~GeV
and very large ($\gtap 200\%$) for $q_T\gtap 250$~GeV.

Our results show that both $q_T$ resummation and finite top-quark mass effects are necessary
to obtain reliable predictions for the double Higgs boson $q_T$ spectrum over the full
transverse momentum range.

\vspace*{1cm}

\subsection*{Acknowledgements}
We would like to thank 
Stefano Catani, Daniel de Florian, Massimiliano Grazzini,  Gudrun Heinrich and Matthias Kerner
for helpful discussions and comments on the manuscript. 
We also thank the authors of Refs.~\cite{NLOHHfullmt,NLOHHfullmt2} for providing us the 
numerical values of the finite part of the two-loop virtual amplitude with full top-quark mass dependence. 
JP would like to thank also Tom Zirke for helpful discussions and Stephan Jahn and Johann Felix von Soden-Fraunhofen for their help
with \texttt{GoSam}.


\newpage

\begin{thebibliography}{99}



\bibitem{Aad:2012tfa}
  G.~Aad {\it et al.} [ATLAS Collaboration],
  Phys.\ Lett.\ B {\bf 716} (2012) 1
  [arXiv:1207.7214 [hep-ex]].

\bibitem{Chatrchyan:2012xdj}
  S.~Chatrchyan {\it et al.} [CMS Collaboration],
  Phys.\ Lett.\ B {\bf 716} (2012) 30
  [arXiv:1207.7235 [hep-ex]].


\bibitem{HCMS}
  V.~Khachatryan {\it et al.} [CMS Collaboration],
  Eur.\ Phys.\ J.\ C {\bf 75} (2015) no.5,  212
  [arXiv:1412.8662 [hep-ex]].

\bibitem{HATLAS}
 G.~Aad {\it et al.} [ATLAS Collaboration],
  Eur.\ Phys.\ J.\ C {\bf 76} (2016) no.1,  6
  [arXiv:1507.04548 [hep-ex]].

\bibitem{Hcombo}
  G.~Aad {\it et al.} [ATLAS and CMS Collaborations],
  JHEP {\bf 1608} (2016) 045
  [arXiv:1606.02266 [hep-ex]].


\bibitem{Aad:2014yja}
  G.~Aad {\it et al.} [ATLAS Collaboration],
  Phys.\ Rev.\ Lett.\  {\bf 114} (2015) no.8,  081802
  [arXiv:1406.5053 [hep-ex]].


\bibitem{Khachatryan:2015yea}
  V.~Khachatryan {\it et al.} [CMS Collaboration],
  Phys.\ Lett.\ B {\bf 749} (2015) 560
  [arXiv:1503.04114 [hep-ex]].

\bibitem{Aad:2015uka}
  G.~Aad {\it et al.} [ATLAS Collaboration],
  Eur.\ Phys.\ J.\ C {\bf 75} (2015) no.9,  412
  [arXiv:1506.00285 [hep-ex]].

\bibitem{Aad:2015xja}
  G.~Aad {\it et al.} [ATLAS Collaboration],
  Phys.\ Rev.\ D {\bf 92} (2015) 092004
  [arXiv:1509.04670 [hep-ex]].


\bibitem{Khachatryan:2016sey}
  V.~Khachatryan {\it et al.} [CMS Collaboration],
  Phys.\ Rev.\ D {\bf 94} (2016) no.5,  052012
  [arXiv:1603.06896 [hep-ex]].


\bibitem{Aaboud:2016xco}
  M.~Aaboud {\it et al.} [ATLAS Collaboration],
  Phys.\ Rev.\ D {\bf 94} (2016) no.5,  052002
  [arXiv:1606.04782 [hep-ex]].


\bibitem{Dolan:2012rv}
  M.~J.~Dolan, C.~Englert and M.~Spannowsky,
  JHEP {\bf 1210} (2012) 112
  [arXiv:1206.5001 [hep-ph]].

\bibitem{Baglio:2012np}
  J.~Baglio, A.~Djouadi, R.~Gr\"ober, M.~M.~M\"uhlleitner, J.~Quevillon and M.~Spira,
  JHEP {\bf 1304} (2013) 151
  [arXiv:1212.5581 [hep-ph]].

\bibitem{Goertz:2013kp}
  F.~Goertz, A.~Papaefstathiou, L.~L.~Yang and J.~Zurita,
  JHEP {\bf 1306} (2013) 016
  [arXiv:1301.3492 [hep-ph]].

\bibitem{Barger:2013jfa}
  V.~Barger, L.~L.~Everett, C.~B.~Jackson and G.~Shaughnessy,
  Phys.\ Lett.\ B {\bf 728} (2014) 433
  [arXiv:1311.2931 [hep-ph]].


\bibitem{deLima:2014dta}
  D.~E.~Ferreira de Lima, A.~Papaefstathiou and M.~Spannowsky,
  JHEP {\bf 1408} (2014) 030
  [arXiv:1404.7139 [hep-ph]]



\bibitem{Dawson:2015oha}
  S.~Dawson, A.~Ismail and I.~Low,
  Phys.\ Rev.\ D {\bf 91} (2015) no.11,  115008
  [arXiv:1504.05596 [hep-ph]].


\bibitem{Eboli:1987dy}
  O.~J.~P.~Eboli, G.~C.~Marques, S.~F.~Novaes and A.~A.~Natale,
  Phys.\ Lett.\ B {\bf 197} (1987) 269.

\bibitem{LOHH1}
  E.~W.~N.~Glover and J.~J.~van der Bij,
  Nucl.\ Phys.\ B {\bf 309} (1988) 282.

\bibitem{LOHH2}
  T.~Plehn, M.~Spira and P.~M.~Zerwas,
  Nucl.\ Phys.\ B {\bf 479} (1996) 46
   Erratum: [Nucl.\ Phys.\ B {\bf 531} (1998) 655]
  [hep-ph/9603205].


\bibitem{NLOHH1}
S.~Dawson, S.~Dittmaier and M.~Spira,
  Phys.\ Rev.\ D {\bf 58} (1998) 115012
  [hep-ph/9805244].

  
 
\bibitem{deFlorian:2013uza}
  D.~de Florian and J.~Mazzitelli,
  Phys.\ Lett.\ B {\bf 724} (2013) 306
  [arXiv:1305.5206 [hep-ph]].

 \bibitem{deflorian-mazzitelli}
   D.~de Florian and J.~Mazzitelli,
  Phys.\ Rev.\ Lett.\  {\bf 111} (2013) 201801
  [arXiv:1309.6594 [hep-ph]].

\bibitem{Grigo:2014jma}
  J.~Grigo, K.~Melnikov and M.~Steinhauser,
  Nucl.\ Phys.\ B {\bf 888} (2014) 17
  [arXiv:1408.2422 [hep-ph]].

 \bibitem{NNLOHH1}
  D.~de Florian, M.~Grazzini, C.~Hanga, S.~Kallweit, J.~M.~Lindert, P.~Maierhöfer, J.~Mazzitelli and D.~Rathlev,
  JHEP {\bf 1609} (2016) 151
  [arXiv:1606.09519 [hep-ph]].



\bibitem{Shao:2013bz}
  D.~Y.~Shao, C.~S.~Li, H.~T.~Li and J.~Wang,
  JHEP {\bf 1307} (2013) 169
  [arXiv:1301.1245 [hep-ph]].

\bibitem{deFlorian:2015moa}
  D.~de Florian and J.~Mazzitelli,
  JHEP {\bf 1509} (2015) 053
  [arXiv:1505.07122 [hep-ph]].


\bibitem{NLOmt1}
  R.~Frederix, S.~Frixione, V.~Hirschi, F.~Maltoni, O.~Mattelaer, P.~Torrielli, E.~Vryonidou and M.~Zaro,
  Phys.\ Lett.\ B {\bf 732} (2014) 142
  [arXiv:1401.7340 [hep-ph]].
  
\bibitem{NLOmt2}
  F.~Maltoni, E.~Vryonidou and M.~Zaro,
  JHEP {\bf 1411} (2014) 079
  [arXiv:1408.6542 [hep-ph]].

\bibitem{Grigo:2013rya}
  J.~Grigo, J.~Hoff, K.~Melnikov and M.~Steinhauser,
  Nucl.\ Phys.\ B {\bf 875} (2013) 1
  [arXiv:1305.7340 [hep-ph]].



 \bibitem{mtexp}
   J.~Grigo, J.~Hoff and M.~Steinhauser,
  Nucl.\ Phys.\ B {\bf 900} (2015) 412
  [arXiv:1508.00909 [hep-ph]].

\bibitem{Degrassi:2016vss}
  G.~Degrassi, P.~P.~Giardino and R.~Gr\"ober,
  Eur.\ Phys.\ J.\ C {\bf 76} (2016) no.7,  411
  [arXiv:1603.00385 [hep-ph]].




\bibitem{NLOHHfullmt}
  S.~Borowka, N.~Greiner, G.~Heinrich, S.~P.~Jones, M.~Kerner, J.~Schlenk, U.~Schubert and T.~Zirke,
  Phys.\ Rev.\ Lett.\  {\bf 117} (2016) no.1,  012001
   Erratum: [Phys.\ Rev.\ Lett.\  {\bf 117} (2016) no.7,  079901]
  [arXiv:1604.06447 [hep-ph]].
 
\bibitem{NLOHHfullmt2}
  S.~Borowka, N.~Greiner, G.~Heinrich, S.~P.~Jones, M.~Kerner, J.~Schlenk and T.~Zirke,
  JHEP {\bf 1610} (2016) 107
  [arXiv:1608.04798 [hep-ph]].



\bibitem{Dokshitzer:hw}
Y.~L.~Dokshitzer, D.~Diakonov and S.~I.~Troian,
Phys.\ Lett.\  B {\bf 79} (1978) 269,
Phys.\ Rep.\ {\bf 58} (1980) 269;
G.~Parisi and R.~Petronzio,
Nucl.\ Phys.\ B {\bf 154} (1979) 427;
%
G.~Curci, M.~Greco and Y.~Srivastava,
Nucl.\ Phys.\ B {\bf 159} (1979) 451.

\bibitem{Collins:1981uk}
J.~C.~Collins and D.~E.~Soper,
Nucl.\ Phys.\ B {\bf 193} (1981) 381
[Erratum-ibid.\ B {\bf 213} (1983) 545];
J.~C.~Collins and D.~E.~Soper,
Nucl.\ Phys.\ B {\bf 197} (1982) 446;
J.~C.~Collins, D.~E.~Soper and G.~Sterman,
Nucl.\ Phys.\ B {\bf 250} (1985) 199.


\bibitem{Kodaira:1981nh}
J.~Kodaira and L.~Trentadue,
Phys.\ Lett.\ B {\bf 112} (1982) 66,
report SLAC-PUB-2934 (1982),
Phys.\ Lett.\ B {\bf 123} (1983) 335.

\bibitem{Catani:2000vq}
S.~Catani, D.~de Florian and M.~Grazzini,
Nucl.\ Phys.\ B {\bf 596} (2001) 299
[hep-ph/0008184].


\bibitem{Bozzi:2005wk}
G.~Bozzi, S.~Catani, D.~de Florian and M.~Grazzini,
Nucl.\ Phys.\ B {\bf 737} (2006) 73
[arXiv:hep-ph/0508068].


\bibitem{Bozzi:2007pn} 
G.~Bozzi, S.~Catani, D.~de Florian and M.~Grazzini,
Nucl.\ Phys.\ B {\bf 791} (2008) 1
[arXiv:0705.3887 [hep-ph]].

\bibitem{Catani:2010pd}
  S.~Catani and M.~Grazzini,
  Nucl.\ Phys.\ B {\bf 845} (2011) 297
  [arXiv:1011.3918 [hep-ph]].

\bibitem{Catani:2013tia}
  S.~Catani, L.~Cieri, D.~de Florian, G.~Ferrera and M.~Grazzini,
  Nucl.\ Phys.\ B {\bf 881} (2014) 414
  [arXiv:1311.1654 [hep-ph]].

\bibitem{Hqt2} 
 D.~de Florian, G.~Ferrera, M.~Grazzini and D.~Tommasini,
  JHEP {\bf 1111} (2011) 064
  doi:10.1007/JHEP11(2011)064
  [arXiv:1109.2109 [hep-ph]].



\bibitem{Catani:2011kr}
  S.~Catani and M.~Grazzini,
  Eur.\ Phys.\ J.\ C {\bf 72} (2012) 2013
   Erratum: [Eur.\ Phys.\ J.\ C {\bf 72} (2012) 2132]
  [arXiv:1106.4652 [hep-ph]].

\bibitem{Catani:2012qa}
  S.~Catani, L.~Cieri, D.~de Florian, G.~Ferrera and M.~Grazzini,
  Eur.\ Phys.\ J.\ C {\bf 72} (2012) 2195
  [arXiv:1209.0158 [hep-ph]].


\bibitem{Bozzi:2010xn}
  G.~Bozzi, S.~Catani, G.~Ferrera, D.~de Florian and M.~Grazzini,
  Phys.\ Lett.\ B {\bf 696} (2011) 207
  [arXiv:1007.2351 [hep-ph]].


\bibitem{pdf1}
 J.~Butterworth {\it et al.},
  J.\ Phys.\ G {\bf 43} (2016) 023001
  [arXiv:1510.03865 [hep-ph]];
%
  S.~Dulat {\it et al.},
  Phys.\ Rev.\ D {\bf 93} (2016) no.3,  033006
  [arXiv:1506.07443 [hep-ph]];
%
  L.~A.~Harland-Lang, A.~D.~Martin, P.~Motylinski and R.~S.~Thorne,
  Eur.\ Phys.\ J.\ C {\bf 75} (2015) no.5,  204
  [arXiv:1412.3989 [hep-ph]];
%
  R.~D.~Ball {\it et al.} [NNPDF Collaboration],
  JHEP {\bf 1504} (2015) 040
  [arXiv:1410.8849 [hep-ph]];
%
  S.~Carrazza, S.~Forte, Z.~Kassabov, J.~I.~Latorre and J.~Rojo,
  Eur.\ Phys.\ J.\ C {\bf 75} (2015) no.8,  369
  [arXiv:1505.06736 [hep-ph]];
%
  J.~Gao and P.~Nadolsky,
  JHEP {\bf 1407} (2014) 035
  [arXiv:1401.0013 [hep-ph]].


\bibitem{Grazzini:2013mca}
  M.~Grazzini and H.~Sargsyan,
  JHEP {\bf 1309} (2013) 129
  [arXiv:1306.4581 [hep-ph]].
  
   
 \bibitem{gosam1}
  G.~Cullen, N.~Greiner, G.~Heinrich, G.~Luisoni, P.~Mastrolia, G.~Ossola, T.~Reiter and F.~Tramontano,
  Eur.\ Phys.\ J.\ C {\bf 72} (2012) 1889
  [arXiv:1111.2034 [hep-ph]].
 \bibitem{gosam2}
  G.~Cullen {\it et al.},
  Eur.\ Phys.\ J.\ C {\bf 74} (2014) no.8,  3001
  [arXiv:1404.7096 [hep-ph]].

 
 \bibitem{cuba1}
  T.~Hahn,
  Comput.\ Phys.\ Commun.\  {\bf 168} (2005) 78
  [hep-ph/0404043].


\bibitem{IImap}
 A.~Daleo, T.~Gehrmann and D.~Maitre,
  JHEP {\bf 0704} (2007) 016
  [hep-ph/0612257].
  
\end{thebibliography}
\end{document}